# "Heading for the Cloud?"
# Implications for Cloud Computing Adopters

**Norah Abokhodair**
Information School, University of Washington, Seattle
noraha@uw.edu

**Hazel Taylor**
Information School, University of Washington, Seattle
hztaylor@uw.edu

**Surry Jones Mowery**
Information School, University of Washington, Seattle
surryjm@gmail.com

**Jitsuko Hasegawa**
Information School, University of Washington, Seattle
jitsukoh@gmail.com

**ABSTRACT**

Cloud computing projects have many implications, including issues such as security, compliance, funding, cohesion with existing systems, operational resource requirements, and number of employees involved. In order to gain a better understanding of why businesses are interested in adopting cloud services in spite of these potential difficulties, we interviewed senior IT personnel at five different organizations about their processes related to "cloud" decisions, their thoughts before and during the process, and the outcome of their endeavor. Our results provide insights from their perspectives into the similarities and differences among the organizations and the implications of "going into the cloud". We conclude with a list of recommended questions and areas to consider for use by other organizations looking into adopting cloud services. The ultimate goal is to help businesses considering a cloud computing project by providing advice from other organizations based on their experience.

**INTRODUCTION**

Information security plays a vital role in protecting the confidentiality, integrity, and availability of an organization's information (Apampa, Wills, and Argles, 2010). The loss of one or more of these attributes can threaten the continued existence of even the largest corporate or government entities. While traditionally, information security has focused on information storage, processing, and transmission within the organization, more recently firms have moved to storing information on external data servers owned and operated by third parties. This is part of a shift toward cloud computing[1] which, with its reliance on third parties for information storage, brings new benefits and risks for an organization's information security.

Cloud computing is a developing domain and its fundamental novelty resides in its characteristics of rapid elasticity for scaling an application when needed, and resource pooling to achieve higher utilization rates, lower costs, and a pay-as-you-go pricing model similar to a utility (Buyya, Broberg, and Goscinski, 2011). Given these benefits, the cloud model has enticed many organizations with promises of lower costs, less physical infrastructure, greater scalability, and agility. However, these potential benefits are accompanied by new risks related to security, data privacy, and compliance. These risks have yet to be fully identified and addressed.

Recent high-profile incidents illustrate how organizations have failed to understand the risks of cloud computing. One such incident involved WikiLeaks, a non-profit organization that used Amazon cloud services to host its content and data. Amazon, on short notice, terminated its services to WikiLeaks and removed the content and data upon request by U.S. federal lawmakers (MacAskill, 2010). Another example is the massive technical failure of Amazon Web Service (AWS) that occurred in April 2011 when Amazon Elastic Compute Cloud went down for several days, despite

---

[1] "Cloud computing is a model for enabling convenient, on-demand network access to a shared pool of configurable computing resources (e.g., networks, servers, storage, applications, and services) that can be rapidly provisioned and released with minimal management effort or service provider interaction." (Mell and Grance, 2010)

Amazon's advertised guarantee of reliability. In addition to disappointing customers, the outage violated one of the main attributes of information security, availability (Mitchell, 2011). These examples demonstrate the vulnerability of organizations that utilize cloud services to host their data and critical resources to decisions by the cloud service provider over which they may have little control (Géczy, Izumi, and Hasida, 2012).

Research into the implications of the adoption of cloud computing is in a nascent stage, and has focused on three primary topics: security, definition and adoption of technology models, and economics. Several studies have examined the implication of the shift to cloud computing on the security of organizations and provided recommendations (see, for example, European Network and Information Security Agency, 2009). Other studies have focused on the definition of cloud computing and explained different adoption models, such as public and private cloud models (Armbrust, Fox, Griffith, Joseph, Katz, Konwinski, Lee, Patterson, Rabkin, Stoica, and Zaharia, 2009). Researchers have also focused on the benefits and economics of adopting cloud computing in large organizations (Harm and Yamartin, 2010). However, little research has been conducted on how organizations deal with and are affected by the aforementioned risks.

Large businesses and government organizations are adopting the cloud rapidly (Stone and Vance, 2010), driven by the significant potential savings in IT costs, and it is likely that medium and smaller firms will also move towards adopting this technology in the near future. A recent study shows that 10% of small businesses have deployed cloud-based systems, while 47% of these businesses are not yet familiar with the cloud approach (Ankeny, 2011). The attraction of cloud computing for smaller organizations is the potential to substantially reduce internal technical support costs so that they can focus on their core businesses (Shacklett, 2011). It is critical for these businesses and their information professionals to gain a better understanding of the various issues and risks associated with cloud computing and how these elements can be managed.

The goal of the present research was to gain a broader understanding of why organizations consider adopting cloud computing, how it might affect their businesses, and how and why they decide to adopt cloud computing or not. Understanding more about the key concerns and experiences of organizations that have already investigated cloud options allows us to provide a preliminary framework that may provide guidance to other organizations facing similar decisions.

## LITERATURE REVIEW

Cloud computing is an enticing new technology, promising benefits of lower cost, less physical infrastructure, greater bandwidth, and ease of use (Jansen and Grance, 2011). The original idea of this technology comes from the economies of scale created by efficiently operated data centers that aggregate massive and various demands for IT resources (Harm and Yamartin, 2010). It is, however, still an immature technology. Security, data privacy, and compliance are major issues that concern security professionals and organizations considering cloud adoption (Jansen and Grance, 2011).

### Deployment models

There are four main deployment models for cloud-based environments: public, private, hybrid, and community cloud. The four models have some features in common: resource distribution, accessibility through networks, and on-demand delivery (Géczy et al., 2012). However, they vary in some unique attributes and the reason they emerged.

The *public cloud* provides utility computing services in a pay-as-you-go manner. Current examples of public cloud utility computing include Amazon Web Services and Microsoft Azure (Armbrust et al., 2009). In contrast, *private cloud* resources and services are owned by an individual organization (Orakwue, 2010) for use by its own members. A private cloud is typically created by restructuring an existing infrastructure to add virtualization and a cloud-like interface. This model allows users within an organization to interact with the local data center and experience the same advantages as users of public clouds, while retaining greater control over data access and security (Buyya et al., 2011).

*Community clouds* arise when a third party provider or the organization itself offers services to a limited set of customers (Mell and Grance, 2010). In these cases, the cloud infrastructure is shared by several organizations and supports a specific community that has shared concerns, such as security requirements or policies (Zissis and Lekkas, 2012).

In *hybrid* or *mixed clouds,* the cloud infrastructure is a composition of two or more clouds (private, community, or public) that remain unique entities but are bound together by standardized or proprietary technology that enables data and application portability (e.g., cloud bursting for load-balancing between clouds) (Mell and Grance, 2010). The hybrid cloud model characterizes a security versus cost compromise between the public and private cloud models.

The organization securely manages core resources and services, and saves costs by outsourcing non-core ones (Géczy et al., 2012).

Deployment models are still under debate. Since the original idea of cloud technology is based on the economies of scale delivered by large-scale data centers, the public cloud model is a logical answer for small businesses seeking to maximize economies of scale. However, because of issues such as regulatory compliance requirements, many cloud vendors started to provide private cloud deployment (and its derivatives). Determining the appropriate deployment model is a key decision point for businesses considering moving to the cloud.

**Benefits of cloud computing**

The National Institute of Standards and Technology suggests five essential benefits of cloud computing for the corporate world (Mell & Grance, 2010), namely on-demand self-service, broad network access, resource pooling, and rapid elasticity.

*On-demand self-service* and *broad network access* offer nearly instant access to resources such as server time and network storage at almost any location, through standard mechanisms that promote use by heterogeneous thin or thick client platforms (e.g., mobile phones, laptops, and PDAs) (Buyya et al., 2011). *Resource pooling* allows the provider to pool computing resources in order to serve multiple consumers using a multi-tenant model, with different physical and virtual resources dynamically assigned and reassigned according to consumer demand.

*Rapid elasticity* provides fast response as an organization's resource requirements expand or contract over time (Armbrust et al., 2009). This is a very attractive benefit for small businesses because they can increase or decrease their resources based on their performance. For example if an small business is making most of its sales during the winter season and less during the summer, then it can have more control over when to add more resources and when to reduce.

**Risks of cloud computing**

The two major risks for cloud computing users are breakdowns in the availability of service and threats to data security (Armbrust et al., 2009). Service breakdowns most commonly occur with a network outage that interrupts user access to the cloud service. Other service concerns include the long-term viability of cloud vendors, and the robustness of the vendors' disaster recovery plans. Additionally, as datacenters grow, they increasingly become the target of distributed denial of service attacks. Data security concerns include threats to data confidentiality and integrity, adequate access control measures, and assurance of complete data disposal at the end of the data life cycle. As data is stored alongside other users' data in the cloud, data segregation with proper encryption is necessary to ensure confidentiality and integrity (Sehgal, Sohoni, Xiong, Fritz, Mulia and Acken., 2011). Users must be well trained in the encryption scheme, because an accident during encryption could make all the data unusable. Access control requires the proper administration of access by in-house legitimate users, and control over data access by privileged vendor administrators. Finally, the practice of achieving data availability in the cloud by duplicated data storage and the synchronization of many virtualized servers makes complete data disposal difficult to verify, once the data is no longer required.

The lack of control over physical entities, as well as the nature of cloud computing, has an impact on an organization's ability to meet regulatory compliance and auditability requirements. Businesses need to carefully examine the trade-offs between the cost savings gained from using public clouds and the controlled compliance benefits available in the private cloud with its higher cost. Additionally, cloud providers operating outside their home country must comply with host country laws, and this introduces foreign legislation risk for clients (Troshani et al., 2011).

Cloud computing clearly provides new opportunities for organizations, and we know that increasing numbers of businesses are trying to take advantage of these. However, there has been little empirical work investigating the factors leading up to business decisions to adopt cloud computing technologies, and we also know little about how firms perceive the benefits and risks of these technologies. The purpose of the current study was to gain a greater understanding of business IT leaders' perceptions and understandings of cloud computing, with a view to providing guidance for organizations considering the shift to this technology.

**METHOD**

Given the limited prior research into the area, we chose an exploratory field study approach for our investigation. We sought a broad understanding of what aspects organizations consider when making decisions about cloud computing options. Our intent was to learn why a variety of organizations with different perspectives and needs were thinking of adopting cloud services, how it would affect their businesses, and how and why they came to their ultimate decision to adopt or not. We contacted senior IT personnel from five organizations, representing a range of different types of business, including non-profit, start-up, government agency, university, and an independent consultant, as shown in Table 1.

## Interviews

A semi-structured interview protocol was developed to guide the interviews. An outline of the interview topics is provided in the first column of Table 2. We wished to explore respondents' knowledge and perspectives on the main issues that have been raised in the literature about cloud computing, in order to identify key drivers and major areas of concern. Respondents were free to answer as little or as much as they wished, and could focus on whichever questions they deemed most relevant to their situation. They were also encouraged to discuss anything else they thought was related to the topic. None of the interviewees answered all of the questions, but the majority of them answered most of the questions.

Three of the authors conducted all interviews, each of which lasted about one hour. Extensive notes were taken during the interviews, and these were integrated across interviewers after each interview. Each question was summarized separately across all five respondents, and cross-interview analysis was used to highlight similarities and differences among the responses. A summary of the interview responses is presented in Table 2.

| Respondent | Organization | Role | Number of IT staff |
|:---:|---|---|:---:|
| A | Independent consultancy | Founder/Owner | - |
| B | Public university | Chief Information Security Officer | 10 |
| C | Government meteorological agency | IT Division Director | 4 direct reports; 280 staff |
| D | Web-based start-up company | Director of Strategic Accounts | 7 |
| E | Museum (non-profit organization) | Senior Systems Engineer | 5 |

**Table 1. Respondent and Organization Details**

## RESULTS

### Interview summaries

In all five cases, the primary motivator for considering cloud computing was cost savings. Secondary motivators included better support for working collaboratively and sharing data; facilities for off-site data back-ups; and a desire to be seen as more energy-efficient and 'green'. The main services being considered were email systems in the cloud and data storage, although migration of operational services to the cloud was also of interest. The expected impact on current IT resource and labor levels varied significantly. While some expected to reduce or maintain current levels, others expected that associated implementation and security issues would require an increase in the level of IT security specialists.

Respondents generally expressed concern about the 'hype' surrounding cloud computing, and suggested that organizations should be skeptical when considering adoption of cloud services. In particular, they suggested that cloud vendors over-simplified the services, making them seem too easy, with the result that businesses might not be fully aware of all the risks involved. In terms of risk, the interviewees identified a number of concerns, mostly revolving around data security, data privacy, and protection, with data and service availability also seen as a critical issue.

## DISCUSSION

These interviews were held with very different types of organizations - arts-based non-profit, government agency, small web-based start-up, large public university and a private consultant in the field. The interviewees and their respective IT staff have almost all been in the industry for at least ten years. Employees within each organization had a range of technical abilities. The non-profit museum employees were considered not "tech-savvy" and their ages ranged from early-twenties to elderly volunteers. The government agency workers were more technically knowledgeable, though they did not know enough about cloud computing before researching the project. The web-based start-up employees were younger (most were in their late twenties to early forties), very comfortable with technology, and eager to learn more. The university had a smaller IT staff and was in need of more resources, considering there are thousands of staff and students relying on their support.

The interviewees were highly motivated, eager, resourceful, and cautious. Most of their research was performed internally, and deliberation about their projects was careful and thoughtful. Each had analyzed the risks involved for his organization and performed some cost-benefit analysis to determine what was appropriate. Some of the organizations had a higher risk level than others. The university, for example, is a higher target in terms of information security attacks, and for its sheer size, visibility, and amount of sensitive data, had a higher amount of risk than the web-based start-up, where no personally identifiable information or financial data was at risk. However, the start-up did face the challenge of abiding by its clients' security rules (Fortune 500 companies), as well as its own. The museum could be at risk for a sensitive data breach with financial data and personally identifiable information, while the government agency was at higher risk for data integrity and service availability. As the agency was responsible for monitoring earthquakes, tsunamis, and other natural disasters, its scientific data required fast and secure transmission, as well as consistent operation of services.

| Respondent / Question | A | B | C | D | E |
|---|---|---|---|---|---|
| Why interested in cloud computing? | Cost savings<br>Distribution of data<br>Virtualization<br>Ability to scale data | Cost savings<br>Most faculty, staff, and students were already using Gmail | Government plan<br>Update cycle<br>Cost savings | Cost savings<br>Fewer required operational resources<br>Off-site backup of data | Letting go of hardware<br>Outsourcing of email<br>Cost, resource savings<br>Very thinly-staffed |
| System structure before using cloud computing? | N/A | Physical data centers | Own the hardware<br>Self-developed software | Physical data centers, internal servers | Internal servers<br>Some virtualized servers |
| Perceived benefits and risks of cloud computing? | Benefit: reputational, saving money, access to equipment beyond the firm's price range<br>Risks: Data privacy, consequences if data integrity is compromised | Risks: Virtual world is not totally secure<br>Unknown percentage of threat<br>Internet is not a stable utility<br>Internet is full of malware | Benefits: Cost savings, Scalability, shorter procurement time<br>Risks: Concern about sharing hardware | Risks: security issues with Fortune 500 clients | Risks: Security issues, risk of leaking personally identifiable information, not within firewall, lack of control |
| Knowledgeability about cloud computing? | Very knowledgeable, but believes the public isn't skeptical enough. | Very knowledgeable | Not enough | Somewhat knowledgeable | Most employees not very "techy". Head IT staff were somewhat knowledgeable, but wanted more information. |
| Did they contract out research? | N/A | Yes (risk assessment team) | Yes | No | No |
| What types of cloud computing and vendors/services? | Common vendors: Microsoft, Amazon, IBM, AT&T.<br>Types: Application, cloud service, infrastructure, platform | Microsoft, Google, Amazon.<br>Public cloud, private cloud for some types of data.<br>Email / collaborative data. | Public cloud<br>Options for self-developing software applications | Storage; secondarily live machines and database (Amazon's EC2).<br>Amazon, based on familiarity and their ability to run a stack | Not sure what they wanted yet - still researching, but interested. |
| Did they adopt cloud computing? Why or why not? | N/A | Partly yes<br>Some data is too sensitive to go on the cloud | Partly yes<br>Some systems are provided for less by cloud<br>Other systems are too specific and cloud doesn't cut cost | Yes: Amazon's S3 storage for backup of site and office data<br>JungleDisk for data encryption<br>Looking into Amazon EC2 – elastic cloud service for disaster recovery | No -- would rather wait until it has been vetted more before jumping in. Wants security, reliability, and redundancy. |

| Was legal advice sought in the research process? | Recommends having legal advice from the very beginning | Yes | No | No | No, but they may be in the future. |
|---|---|---|---|---|---|
| How satisfied with provider? Why or why not? | N/A | Satisfied | N/A | Satisfied with Amazon Unhappy with separate application based on cloud hosting - outages | N/A |
| Actual value/risk since adopting? | How would you know if there was a breach? How would you know where the data is going? Does the provider know where the data is and where it is going? | Risk: Organization is a major target, and there will be those who wish to break into the cloud. Risk of not knowing when and how bad it will be. | N/A | Value: Cost and resources Risk: Data privacy issues, additional security reviews for clients | N/A |
| What is the quality of customer service? | N/A | Good | N/A | Does not know | N/A |
| Do they have any regrets or things they wished they had done, post-decision? | N/A | Worried about attacks against virtual environments | N/A | No | N/A |
| Has cost been on budget? | N/A | Not yet | N/A | Yes | N/A |
| Do you have any advice for people who are looking into cloud services? | Consider total cost of ownership. Develop plan for change. Look at the business holistically -- what is the business element with security? Use checklists from Cloud Security Alliance (2009) document. | Be informed. Choose established providers. Cost benefit analysis. Seek reference sites from potential providers. | N/A | Do your research. Calculate the cost for the organization. Do financial modeling to ensure it makes sense. Take security in your own hands for the product transfer. | More interested in virtualization than cloud computing. Other non-profits are using it. Able to stay within firewall, use for storage and redundancy. Able to shift loads more easily. |

**Table 2. Summarized Interview Questions and Responses**

The nature of the organizations also had an effect on the decisions of the IT staff. The start-up is a very small company with a horizontal structure, and decisions were easier and quicker to make. The museum had experienced several rounds of layoffs and budgets were tight: many projects had been put on hold until more money came in. The government agency is bound by strict rules and regulations that caused projects to take longer to complete because of the need to meet all regulatory demands. As a non-profit public institution serving thousands of students and staff every year, the university is a highly-visible target with a large amount of sensitive data to protect, but has to rely on government funding, which fluctuates according to other conditions.

Despite these differences between organizations, there were similarities in their decision processes regarding the adoption of cloud services. Cost savings, scalability, fewer required operational resources, and increased computing power were all major reasons for considering cloud services. IT staff numbers generally declined during the 2009 economic downturn and organizations are looking into doing more for less. Outsourcing email through the cloud was the main driver for the university and the museum, while the start-up looked into data storage and backups, and the government agency looked into a platform-as-a-service solution for self-developed software applications. The museum, government agency and start-up were all considering a public cloud service, while the university was considering a combination of public and private cloud. Security was clearly the biggest concern for all the organizations, requiring detailed research to ensure that the cloud was the right choice. The level of concern regarding security, however, differed for each organization. While the start-up decision makers were concerned with the security issues for each client, they were confident in their choice to use cloud service and took extra steps for data encryption. The lack of sensitive data involved likely contributed to their approach on security, as well as the fact that they were a small, growing start-up company trying to stay afloat: the employees took on multiple roles and were extremely busy with many other areas to consider.

The university went ahead, after a thorough risk analysis, with the decision to adopt cloud service despite heavy security issues. Its contract with a major cloud service provider was also heavily negotiated to help produce a better compromise. In addition, it was selective about what data was moved into the cloud, and some information was not migrated due to security demands that could not be fully addressed.

The government agency also partially adopted cloud service. Its operational stability concerns were not fully addressed by the cloud providers, and a smaller compromise was established. Security concerns were very high with the museum, and its ultimate decision was to "wait and see" how the industry evolves over time. With limited funding, the museum is very cautious with approval of projects and higher priority demands must be addressed first.

One of the most interesting aspects of the interviews was the advice that the interviewees had for those looking into cloud services, as summarized in Table 3. While performing extensive research and a cost-benefit analysis were all recommended, each respondent contributed different and valuable advice. The independent consultant recommended the Cloud Security Alliance white paper (2009), and also reminded organizations to consider the total cost of ownership, as well as the cost of change. His work as a private IT consultant likely contributed to his perspective, particularly his advice to take a holistic view of security and the business, and not to view each in a vacuum.

The university respondent recommended that organizations choose established providers when researching cloud services, and ask providers for client reference sites to gain insight into their experience. The main advice from the start-up company was a reminder to take a proactive approach to security: if certain security measures are not offered from a vendor, look for other vendors or consider combinations of products and vendors. Lastly, the museum reminded other organizations (in particular, non-profits) to research virtualization as a way to make more efficient use of in-house computing resources if there are security or cost issues with cloud computing options.

## CONCLUSION

While most of our findings echo the advice found in industry best practice recommendations, our findings do provide a greater level of detail on specific considerations that organizations should keep in mind when evaluating cloud computing options. The list of recommendations shown in Table 3 incorporates best practice from the literature and the advice and feedback of professionals in the field, drawn from their work experience. The professional viewpoint obtained from our respondents adds breadth and depth in terms of detail and other areas and questions that organizations should consider regarding a possible cloud-computing endeavor. In sum, while cloud computing is a new and rapidly changing technology, there is already a lot of information available to organizations considering adopting this technology. As with all new technologies, organizations are advised to approach cloud opportunities with some skepticism, and base their decisions on thorough research and a careful consideration of the organization's needs.

| Area | Recommendation |
|---|---|
| Research | Do as much research as possible. Information is available regarding cloud computing, including the Cloud Security Alliance (2009) white paper and NIST documents (Mell and Grance 2010, Jansen and Grance 2011). |
| Needs analysis | Perform a detailed analysis of organizational needs, strategic objectives and business goals, and determine the available IT options that will support the goals. Identify key stakeholders within the organization. Identify vendors and their offerings. Consider the impact of proposed solutions on current systems. Determine implementation and training requirements. |
| Reference sites | Look to other similar organizations for information and advice about their experiences with cloud services and the vendors they are using. Obtain reference sites from short-listed vendors and evaluate their experiences. |
| Risk analysis | Determine risks, security, and compliance issues, including any client security practices that the organization must address. Research risk mitigation options beyond what the provider offers; for example, data encryption. |
| Cost-benefit analysis | Perform a cost-benefit analysis and evaluate all alternatives, including non-cloud options. Consider the total cost of ownership and the cost of change for each option. |
| Legal implications | Consult the organization's legal team on security and compliance issues. Ensure that the legal team vets all provider contracts and service-level agreements. |
| Organizational impact | Consider the appropriate decision for the organization, keeping in mind the industry, culture, type of organization, characteristics of employees, and risk propensity. Seek the decision that best fits the organization holistically, from business and technical perspectives. Consider the impact on organizational staff in general, and IT staff in particular. Determine the behavioral changes that will be required of staff, and consider what support will be required and who will provide it. |

**Table 3. Summary of Recommendations for Assessing Cloud Services**